\documentclass{amsart}
\usepackage{amssymb}
\usepackage{amsmath}

 \theoremstyle{plain}
\newtheorem{theorem}{Theorem}[section]
\newtheorem{lemma}[theorem]{Lemma}
\newtheorem{proposition}[theorem]{Proposition}
\theoremstyle{definition}
\newtheorem{definition}[theorem]{Definition}

\theoremstyle{remark}

\numberwithin{equation}{section}

\input{tcilatex}

\begin{document}
\title[Delsarte operators in parametric spaces]{THE GENERAL
DIFFERENTIAL-GEOMETRIC STRUCTURE OF MULTIDIMENSIONAL DELSARTE TRANSMUTATION
OPERATORS IN PARAMETRIC FUNCTIONAL SPACES AND THEIR APPLICATIONS IN SOLITON
THEORY. Part 2}
\author{J. Golenia*)}
\address{The AMM University of Science and Technology, Department of Applied
Mathematics, Krakow 30059 Poland}
\email{jnapora@wms.mat.agh.edu.pl}
\thanks{The fourth author was supported in part by a local AGH grant.}
\author{Y.A. Prykarpatsky*)**), A.M. Samoilenko**)}
\address{Intitute of Mathematics at the NAS, Kiev 01601, Ukraine, and the
AMM University of Science and Technology, Department of Applied Mathematics,
Krakow 30059 Poland}
\curraddr{Brookhaven Nat. Lab., CDIC, Upton, NY, 11973 USA}
\email{yarpry@bnl.gov, sam@imath.kiev.ua}
\author{A.K.~Prykarpatsky*)***)}
\address{The AMM University of Science and Technology, Department of Applied
Mathematics, Krakow 30059 Poland, and Dept. of \ Nonlinear Mathematical
Analysis at IAPMM, NAS of Ukraine, Lviv 79601 Ukraina}
\email{prykanat@cybergal.com, pryk.anat@ua.fm}
\dedicatory{This paper is dedicated to the memory of 95 -th birthday and
10-th death anniversaries of the mathematics and physics luminary of the
former century academician \textbf{Nikolay Nikolayevich Bogoliubov} }
\date{}

\begin{abstract}
The structure properties of multidimensional Delsarte transmutation
operators in parametirc functional spaces are studied by means of
differential-geometric tools. It is shown that kernels of the corresponding
integral operator expressions depend on the topological structure of related
homological cycles in the coordinate space. As a natural realization of the
construction presented we build pairs of Lax type commutive differential
operator expressions related via a Darboux-Backlund transformation having a
lot of applications in solition theory. Some results are also sketched
concerning theory of \ Delsarte transmutation operators for affine
polynomial pencils of \ multidimensional differential operators.
\end{abstract}

\maketitle


\section{Introduction}

\setcounter{equation}{0}Consider the Banach space ${\mathcal{H}}%
:=L_{2}(l;H), $ $H:=L_{2}(\mathbb{R}^{m};\mathbb{C}^{N}),$ with the natural
semi-linear scalar form on ${\mathcal{H}}^{\ast }\times {\mathcal{H}}$:

\begin{equation}
(\varphi ,\psi )_{\mathcal{H}}:=\int_{l}dt\int_{\mathbb{R}^{m}}dx\bar{\varphi%
}^{\intercal }(t;x)\psi (t;x),  \label{1.1}
\end{equation}%
where $(\varphi ,\psi )\in \mathcal{H\times }\mathcal{H},$ $t\in l:=[0,T)\in
\mathbb{R}_{+}$ is an evolution parameter, $N\in \mathbb{Z}_{+},$ $"-"$ is
the complex conjugation and the sign $"\intercal "$ means the usual matrix
transposition. Take now a pair of closed dense subspaces $\mathcal{H}_{0}$
and $\tilde{{\mathcal{H}}_{0}}$ in $\mathcal{H}$ and two linear differential
operators of equal order $\mathcal{L}:=\frac{\partial }{\partial t}-L$ \ and
$\mathcal{\tilde{L}}:=\frac{\partial }{\partial t}-\tilde{L}$ \ from $%
\mathcal{H}$ to $\mathcal{H},$ where

\begin{equation}
L:=\sum_{|\alpha |=0}^{n(L)}a_{\alpha }(t;x)\frac{\partial ^{|\alpha |}}{%
\partial x^{\alpha }},\text{ }\widetilde{L}:=\sum_{|\beta |=0}^{n(\widetilde{%
L})}\tilde{a}_{\beta }(t;x)\frac{\partial ^{|\beta |}}{\partial x^{\beta }},
\label{1.2}
\end{equation}%
$(t;x)\in l\times \mathbb{R}^{m}$ and coefficients $a_{\alpha },\tilde{a}%
_{\beta }\in C^{1}(\mathbb{R};S(\mathbb{R}^{m};End\mathbb{C}^{N}))$ for all $%
|\alpha |,|\beta |=\overline{0,n},$ $\ n(L):=n=:n(\tilde{L}).$

\begin{definition}
(J. Delsarte and J. Lions [2]). A linear invertible operator
$\bold{\Omega}$ defined on the whole $ \mathcal{H}$ and acting
from  ${\mathcal{H}}_{0}$ onto $ \tilde{\mathcal{H}}$ is called a
Delsarte transmutation operator for the pair of differential
operators $L$ and $\tilde{L}$, if the following two conditions
hold:

$\bullet$ the operator $\bold{\Omega}$ and its inverse $\bold{\Omega}^{-1}$ are continuous in
$\mathcal{H}$;

$\bullet$ the operator identity $ \bold{\Omega}L=\tilde{L} {\bold\Omega}$ is satisfied.
\end{definition}

Such transmutation operators were for the first time introduced in [1,2] for
the case of one-dimensional second order differential operators. In
particular, for the Sturm-Liouville and Dirac operators the complete
structure of the corresponding Delsarte transmutation operators was
described in [3,4], where also the extensive applications to spectral theory
were given. A special generalization of the Delsarte transmutation operator
for two-dimensional Dirac operators was done for the first type in [5],
where its applications to inverse scattering theory and solving some
nonlinear two-dimensional evolution equations were also presented.

Recently some progress in this direction was made in [6,7] due to analyzing
a special operator structure of Darboux type transformation which appeared
in [8]. In this work we describe the general differential-geometric and
topological structure of multi-dimensional Delsarte type transmutation
operators for differential expressions like (1.2) acting in parametric
functional spaces, by means of the differential-geometric approach devised
in [6,7] and discuss some of their applications to Darboux-Backlund
transformations and soliton theory.

\section{The differential-geometric structure of the generalized Lagrangian
identity}

Take a multi-dimensional differential operator $\mathcal{L}:=L-\partial
/\partial t:\mathcal{H}\longrightarrow \mathcal{H}$ given above and write
down its formally adjoint expression as
\begin{equation}
\mathcal{L}^{\ast }(t;x|\partial ):=\sum_{|\alpha |=0}^{n(L)}(-1)^{|\alpha |}%
\frac{\partial ^{|\alpha |}}{\partial x^{\alpha }}\cdot \bar{a}_{\alpha
}^{\intercal }(t;x)+\partial /\partial t  \label{2.1}
\end{equation}%
with $(t;x)\in l\times \mathbb{R}^{m}.$ Consider the following easily
derivable generalized Lagrangian identity in :
\begin{equation}
\begin{array}{l}
<\varphi ,\mathcal{L}\psi >-<\mathcal{L}^{\ast }\varphi ,\psi > \\
=\sum_{i=1}^{m}(-1)^{i+1}\frac{\partial }{\partial x_{i}}Z_{i}[\varphi ,\psi
]-\frac{\partial }{\partial t}(\bar{\varphi}^{\intercal }(x)\psi (x))%
\end{array}
\label{2.2}
\end{equation}%
where for any pair $(\varphi ,\psi )\in D(\mathcal{L}^{\ast })\times D(%
\mathcal{L})$ \ from a dense domain $D(\mathcal{L}^{\ast })\times D(\mathcal{%
L})\subset \mathcal{H}^{\ast }\times \mathcal{H}$ \ the mappings $%
Z_{i}[\varphi ,\psi ]:\mathcal{H}^{\ast }\times \mathcal{H\rightarrow }%
\mathbb{C},$ $i=\overline{1,m},$ are semilinear$\ $ for each $(t;x)\in
l\times \mathbb{R}^{m}.$

The Lagrangian expression (2.2) can be analyzed effectively by means of the
following differential-geometric construction: having multiplied (2.2) by
the oriented Lebesgue measure $dt\wedge dx,$ $dx:=(\underset{j=%
\overrightarrow{1,m}}{\wedge }dx_{j}),$ we easily obtain that
\begin{equation}
(<\varphi ,L\psi -\frac{\partial \varphi }{\partial t}>-<L^{\ast }\varphi +%
\frac{\partial \varphi }{\partial t},\psi >)dt\wedge dx=dZ^{(m)}[\varphi
,\psi ],  \label{2.3}
\end{equation}%
where $Z^{(m)}[\varphi ,\psi ]\in \Lambda ^{m}(\mathbb{R}^{1+m};\mathbb{C})$
is a differential $m$-form on $\mathbb{R}\times \mathbb{R}^{m}$ given by the
expression
\begin{equation}
\begin{array}{l}
Z^{(m)}[\varphi ,\psi ]=\sum_{\overline{1,m}}dx_{1}\wedge dx_{2}\wedge
...\wedge dx_{i-1}\wedge Z_{i}[\varphi ,\psi ]dx_{i+1}\wedge ...\wedge dx_{m}
\\
-\bar{\varphi}^{\intercal }(x;t)\psi (x;t)dx.%
\end{array}
\label{2.4}
\end{equation}%
Take now a pair $(\varphi (\lambda ),\psi (\mu ))\in \mathcal{H}%
_{0}^{\circledast }\times \mathcal{H}_{0}$ with $\lambda ,\mu \in \Sigma ,$
where $\ \Sigma \subset \mathbb{C}$ is some "spectral" space of paramers, $%
\mathcal{H}_{0}^{\circledast }$ and $\mathcal{H}_{0}\subset \mathcal{H}$ are
the corresponding closed subspaces of $\mathcal{H}^{\ast }$ and $\mathcal{H}%
, $ being defined as solutions to the following evolution equations:
\begin{equation}
\partial \psi /\partial t=L\psi ,\qquad \partial \varphi /\partial
t=-L^{\ast }\varphi  \label{2.5}
\end{equation}%
with Cauchy data $\psi |_{t=t_{0}}=\bar{\psi}_{\lambda }\in L_{2}(\mathbb{R}%
^{m};\mathbb{C}^{N})$ and $\varphi |_{t=t_{0}}=\bar{\varphi}_{\mu }\in L_{2}(%
\mathbb{R}^{m};\mathbb{C}^{N})$ for $\lambda ,\mu \in \Sigma $ at $t_{0}\in
l,$ being fixed, $\psi |_{\Gamma }=0$ and $\varphi |_{\Gamma }=0$ for some
chosen piece-wise smooth hypersurface $\Gamma $ \ $\subset \mathbb{R}^{m}.$

Having assumed that linear differential equations (2.5) are solvable for all
$t\in \lbrack t_{0},T),$ $t_{0}<T\in \mathbb{R}_{+},$ we can obtain right
away from (2.3) and (2.4) that the differential $m$-form $Z^{(m)}[\varphi
,\psi ](\eta |\xi )\in \Lambda ^{m}(\mathbb{R}^{1+m};\mathbb{C})$ is closed
for any $\eta ,\xi \in \Sigma $ \ and $(\varphi ,\psi )\in \mathcal{H}%
_{0}^{\circledast }\times \mathcal{H}_{0}.$ Thereby, due to the well known
Poincare lemma [12] one can state that there exists an $(m-1)$-differential
form $\Omega ^{(m-1)}[\varphi ,\psi ](\eta |\xi )\in \Lambda ^{m-1}(\mathbb{R%
}^{1+m};\mathbb{C})$ satisfying the equality
\begin{equation}
Z^{(m)}[\varphi ,\psi ](\eta |\xi )=d\Omega ^{(m-1)}[\varphi ,\psi ](\eta
|\xi )  \label{2.6}
\end{equation}%
since $d^{2}\equiv 0$ on the Grassmann algebra $\Lambda (\mathbb{R}^{1+m};%
\mathbb{C})$ of differential forms on $\mathbb{R}\times \mathbb{R}^{m}.$
Take now an arbitrary $m$-dimensional piecewise smooth hyper-surface $%
S(\sigma _{(x,t)}^{(m-1)},\sigma _{(x_{0},t_{0})}^{(m-1)})$ $\subset \mathbb{%
R}\times \mathbb{R}^{m}$ spanning some two $(m-1)$-dimensional
homological cycles $\sigma _{(x,t)}^{(m-1)},$ $\sigma
_{(x_{0},t_{0})}^{(m-1)}$ marked by points $(x,t)$ and
$(x_{0},t_{0})\in \mathbb{R}^{m}\times l,$ in such a way that
$$\partial S(\sigma _{(x,t)}^{(m-1)},\sigma
_{(x_{0},t_{0})}^{(m-1)})=\sigma _{(x,t)}^{(m-1)}-\sigma
_{(x_{0},t_{0})}^{(m-1)}$$ and related in some way with the chosen
above hypersurface $\Gamma $ \ $\subset \mathbb{R}^{m}.$ Then one
gets from (2.6) that due to the Stokes theorem [12]
\begin{equation}
\begin{array}{l}
\int_{{\mathcal{S}}(\sigma _{(x,t)}^{(m-1)},\sigma
_{(x_{0},t_{0})}^{(m-1)})}Z^{(m)}[\varphi ,\psi ](\eta |\xi ) \\
=\int_{\sigma _{(x,t)}^{(m-1)}}\Omega ^{(m-1)}[\varphi ,\psi ](\eta |\xi
)-\int_{\sigma _{(x_{0},t_{0})}^{(m-1)}}\Omega ^{(m-1)}[\varphi ,\psi ](\eta
|\xi ) \\
:=\Omega _{(x,t)}[\varphi ,\psi ](\eta |\xi )-\Omega
_{(x_{0},t_{0})}[\varphi ,\psi ](\eta |\xi ),%
\end{array}
\label{2.7}
\end{equation}%
\begin{equation*}
\begin{array}{l}
\int_{{\mathcal{S}}(\sigma _{(x,t)}^{(m-1)},\sigma _{(x_{0},t_{0})}^{(m-1)})}%
\bar{Z}^{(m),\intercal }[\varphi ,\psi ](\eta |\xi ) \\
=\int_{\sigma _{(x,t)}^{(m-1)}}\bar{\Omega}^{(m-1),\intercal }[\varphi ,\psi
](\eta |\xi )-\int_{\sigma _{(x_{0},t_{0})}^{(m-1)}}\bar{\Omega}%
^{(m-1),\intercal }[\varphi ,\psi ](\eta |\xi ) \\
:=\Omega _{(x,t)}^{\circledast }[\varphi ,\psi ](\eta |\xi )-\Omega
_{(x_{0},t_{0})}^{\circledast }[\varphi ,\psi ](\eta |\xi ),%
\end{array}%
\end{equation*}%
where the expressions
\begin{equation*}
\Omega _{(x,t)}[\varphi ,\psi ](\eta |\xi ),\text{\ }\Omega
_{(x_{0},t_{0})}[\varphi ,\psi ](\eta |\xi ),\text{ }\Omega
_{(x,t)}^{\circledast }[\varphi ,\psi ](\eta |\xi ),\text{ }\Omega
_{(x_{0},t_{0})}^{\circledast }[\varphi ,\psi ](\eta |\xi )
\end{equation*}%
with $\eta ,\xi \in \Sigma $ are also considered as the corresponding
kernels of invertible integral operators\ $\Omega _{(x,t)}[\varphi ,\psi ],$
$\Omega _{(x_{0},t_{0})}[\varphi ,\psi ],$ $\Omega _{(x,t)}^{\circledast
}[\varphi ,\psi ],$ $\Omega _{(x_{0},t_{0})}^{\circledast }[\varphi ,\psi ]$
in $L_{2}^{(\rho )}(\Sigma ;\mathbb{C})$ of measured functions on $\Sigma $
with respect to a finite Borel measure $\rho $ on Borel subsets from $\Sigma
$ \ for any $(x,t)\in \mathbb{R}^{m}\times l,$ considered here as parameters$%
.$ Moreover, the homotopy conditions in the space $L_{2}^{(\rho )}(\Sigma ;%
\mathbb{C})$
\begin{equation}
\lim_{(x,t)\rightarrow (x_{0},t_{0})}\Omega _{(x,t)}[\varphi ,\psi ]=\Omega
_{(x_{0},t_{0})}[\varphi ,\psi ],\text{ \ }\lim_{(x,t)\rightarrow
(x_{0},t_{0})}\Omega _{(x,t)}^{\circledast }[\varphi ,\psi ]=\Omega
_{(x_{0},t_{0})}^{\circledast }[\varphi ,\psi ]  \label{2.8}
\end{equation}%
are assumed to be satisfied for all $(\varphi ,\psi )\in {\mathcal{H}}%
_{0}^{\circledast }\times {\mathcal{H}}_{0}.$

\section{The multidimensional Delsarte transmutation operators and their $m$%
-dimensional topological structure}

For a Delsarte transmutation operators $\mathbf{\Omega }:{\mathcal{H}}%
\rightarrow {\mathcal{H}}$ and $\mathbf{\Omega }^{\circledast }:{\mathcal{H}}%
^{\ast }\rightarrow {\mathcal{H}}^{\ast }$ to be constructed ab initio, it
is necessary in accordance with Def. 1.1 to define the corresponding closed
two subspaces $\tilde{\mathcal{H}}_{0}\subset {\mathcal{H}}$ and $\tilde{%
\mathcal{H}}_{0}^{\circledast }\subset {\mathcal{H}}^{\ast }.$

Let now
\begin{equation}
\begin{array}{c}
\tilde{\mathcal{H}}_{0}:=\{\tilde{\psi}\in {\mathcal{H}}_{0}:\tilde{\psi}%
=\psi \Omega _{(x,t)}^{-1}[\varphi ,\psi ]\Omega _{(x_{0},t_{0})}[\varphi
,\psi ],\text{ \ \ \ \ }\tilde{\psi}|_{\tilde{\Gamma}}=0\}, \\
\tilde{\mathcal{H}}_{0}^{\circledast }:=\{\tilde{\varphi}\in {\mathcal{H}}%
_{0}^{\ast }:\tilde{\varphi}=\varphi (\Omega _{(x,t)}^{\ast }[\varphi ,\psi
])^{-1}\Omega _{(x_{0},t_{0})}^{\ast }[\varphi ,\psi ],\text{ }\tilde{\varphi%
}|_{\tilde{\Gamma}}=0\}%
\end{array}
\label{3.1}
\end{equation}%
for some hypersurface $\tilde{\Gamma}$ $\subset \mathbb{R}^{m}$ related in
some way with hypersurfaces $\Gamma $ and $\Gamma ^{\ast }$ chosen before,
where the operators $\Omega _{(x,t)}^{-1}[\varphi ,\psi ],(\Omega
_{(x,t)}^{\circledast }[\varphi ,\psi ])^{-1}:L_{2}^{(\rho )}(\Sigma ;%
\mathbb{C})\longrightarrow L_{2}^{(\rho )}(\Sigma ;\mathbb{C})$ are
correspondingly inverse to the scalar operators $\Omega _{(x,t)}[\varphi
,\psi ],$ $\Omega ^{\circledast }[\varphi ,\psi ]:L_{2}^{(\rho )}(\Sigma ;%
\mathbb{C})\longrightarrow L_{2}^{(\rho )}(\Sigma ;\mathbb{C}),$ \
parametrized by variables $(x,t)\in \mathbb{R}^{m}\times l.$ Due to the
properties of operators%
\begin{equation*}
\Omega _{(x,t)}[\varphi ,\psi ],\text{ }\Omega _{(x_{0},t_{0})}[\varphi
,\psi ],\text{ }\Omega _{(x,t)}^{\circledast }[\varphi ,\psi ],\text{ }%
\Omega _{(x_{0},t_{0})}^{\circledast }[\varphi ,\psi ]
\end{equation*}%
in the space $L_{2}^{(\rho )}(\Sigma ;\mathbb{C}),$ \ the spaces (3.1) are
also closed in $\mathcal{H}$ and ${\mathcal{H}}^{\circledast },$
correspondingly. Expressions (3.1) define the following actions
\begin{equation}
\mathbf{\Omega }:\psi \rightarrow \tilde{\psi},\qquad \mathbf{\Omega }%
^{\circledast }:\varphi \rightarrow \tilde{\varphi}  \label{3.2}
\end{equation}%
for any arbitrary but fixed (!) pair of functions $(\varphi ,\psi )\in {%
\mathcal{H}}_{0}^{\circledast }\times {\mathcal{H}}_{0}.$ For retrieving
these actions upon the whole space ${\mathcal{H}}^{\ast }\times {\mathcal{H}}
$ at a fixed pair of functions $(\varphi ,\psi )\in {\mathcal{H}}%
_{0}^{\circledast }\times {\mathcal{H}}_{0},$ let us make use of the well
known method of variation of constant:
\begin{equation}
\begin{array}{l}
\mathbf{\Omega }\cdot \psi :=\tilde{\psi} \\
=\psi \Omega _{(x,t)}^{-1}[\varphi ,\psi ](-\int_{{\mathcal{S}}(\sigma
_{(x,t)}^{(m-1)},\sigma _{(x_{0},t_{0})}^{(m-1)})}Z^{(m)}[\varphi ,\psi
]+\Omega _{(x,t)}[\varphi ,\psi ]) \\
=\psi -\psi \Omega _{(x,t)}^{-1}[\varphi ,\psi ]\Omega
_{(x_{0},t_{0})}[\varphi ,\psi ]\Omega _{(x_{0},t_{0})}^{-1}[\varphi ,\psi
]\int_{{\mathcal{S}}(\sigma _{(x,t)}^{(m-1)},\sigma
_{(x_{0},t_{0})}^{(m-1)})}Z^{(m)}[\varphi ,\psi ] \\
=\psi -\tilde{\psi}\Omega _{(x_{0},t_{0})}^{-1}[\varphi ,\psi ]\int_{{%
\mathcal{S}}(\sigma _{(x,t)}^{(m-1)},\sigma
_{(x_{0},t_{0})}^{(m-1)})}Z^{(m)}[\varphi ,\psi ] \\
=(1-\tilde{\psi}\Omega _{(x_{0},t_{0})}^{-1}[\varphi ,\psi ]\int_{{\mathcal{S%
}}(\sigma _{(x,t)}^{(m-1)},\sigma _{(x_{0},t_{0})}^{(m-1)})}Z^{(m)}[\varphi
,\cdot ])\psi ; \\
\hat{\Omega}^{\circledast }\cdot \varphi :=\tilde{\varphi} \\
=\varphi (\Omega _{(x,t)}^{\circledast }[\varphi ,\psi ])^{-1}(-\int_{{%
\mathcal{S}}(\sigma _{(x,t)}^{(m-1)},\sigma _{(x_{0},t_{0})}^{(m-1)})}\bar{Z}%
^{(m),\intercal }[\varphi ,\psi ]+\Omega _{(x,t)}^{\circledast }[\varphi
,\psi ]) \\
=\varphi -\varphi (\Omega _{(x,t)}^{\circledast }[\varphi ,\psi
])^{-1}\Omega _{(x_{0},t_{0})}^{\circledast }[\varphi ,\psi ](\Omega
_{(x_{0},t_{0})}^{\circledast }[\varphi ,\psi ])^{-1}\int_{{\mathcal{S}}%
(\sigma _{(x,t)}^{(m-1)},\sigma _{(x_{0},t_{0})}^{(m-1)})}\bar{Z}%
^{(m),\intercal }[\varphi ,\psi ] \\
=(1-\tilde{\varphi}(\Omega _{(x_{0},t_{0})}^{\circledast }[\varphi ,\psi
])^{-1}\int_{{\mathcal{S}}(\sigma _{(x,t)}^{(m-1)},\sigma
_{(x_{0},t_{0})}^{(m-1)}))}\bar{Z}^{(m),\intercal }[\cdot ,\psi ])\varphi ,%
\end{array}
\label{3.3}
\end{equation}%
where $(\varphi ,\psi )\in {\mathcal{H}}_{0}^{\ast }\times {\mathcal{H}}_{0}$
and parameters $(x,t)\in \mathbb{R}^{m}\times (t_{0},T)$ are arbitrary.
Thereby, due to (3.3) one can define invertible extended Delsarte
transmutation operators
\begin{equation}
\begin{array}{l}
\mathbf{\Omega }:=1-\tilde{\psi}(\Omega _{(x_{0},t_{0})}[\varphi ,\psi
])^{-1}\int_{{\mathcal{S}}(\sigma _{(x,t)}^{(m-1)},\sigma
_{(x_{0},t_{0})}^{(m-1)})}Z^{(m)}[\varphi ,\cdot ], \\
\mathbf{\Omega }^{\circledast }:=1-\tilde{\varphi}(\Omega
_{(x_{0},t_{0})}^{\ast }[\varphi ,\psi ])^{-1}\int_{{\mathcal{S}}(\sigma
_{(x,t)}^{(m-1)},\sigma _{(x_{0},t_{0})}^{(m-1)})}\bar{Z}^{(m),\intercal
}[\cdot ,\psi ],%
\end{array}
\label{3.4}
\end{equation}%
acting, correspondingly, ithe whole spaces $\mathcal{H}$ and ${\mathcal{H}}%
^{\ast }.$

Consider now the following commutative diagram
\begin{equation*}
\begin{array}{ccc}
{\mathcal{H}} & \overset{\frac{\partial }{\partial t}-L}{\rightarrow } & {%
\mathcal{H}} \\
\mathbf{\Omega }\downarrow &  & \downarrow \mathbf{\Omega } \\
{\mathcal{H}} & \overset{\frac{\partial }{\partial t}-\tilde{L}}{\rightarrow
} & {\mathcal{H}},%
\end{array}%
\end{equation*}%
which defines the transformed operator $(\tilde{L}-\frac{\partial }{\partial
t}):{\mathcal{H}}\longrightarrow {\mathcal{H}}$ by means of the Delsarte
transmutation expression $\frac{\partial }{\partial t}-\tilde{L}=\mathbf{%
\Omega }(\frac{\partial }{\partial t}-L)\mathbf{\Omega }^{-1}.$The pair of
functions $(\tilde{\varphi},\tilde{\psi})\in \tilde{\mathcal{H}}%
_{0}^{\circledast }\times \tilde{\mathcal{H}}_{0}$ and the operator (3.5)
are described by the following proposition.

\begin{proposition}
The pair of transformed functions
$(\tilde{\varphi},\tilde{\psi}) \in \tilde{\mathcal{H}}_{0}^{?} \times \tilde{\mathcal{H}}_{0}$
solves, correspondingly, the evolution equations
\begin{equation}
\partial \tilde{\psi}/\partial t=\tilde{L}\tilde{\psi},\qquad
\partial \tilde{\varphi}/\partial t=-\tilde{L}^{*}\tilde{\varphi}
\end{equation}
for all $t \in (t_{0},T).$
\end{proposition}

\begin{proof}
It is enough to consider for any $\psi \in \tilde{{\mathcal{H}}}_{0}$ the
expressions
\begin{equation*}
(\frac{\partial }{\partial t}-\tilde{L})\tilde{\psi}=\mathbf{\Omega }(\frac{%
\partial }{\partial t}-L)\mathbf{\Omega }^{-1}\tilde{\psi}=\mathbf{\Omega }(%
\frac{\partial }{\partial t}-L)\psi =0
\end{equation*}%
which holds due to the definition of \ the closed subspace ${\mathcal{H}}%
_{0}.$ The equality \ $(\partial \tilde{\varphi}/\partial t$ $+\tilde{L}%
^{\ast })\tilde{\varphi}=0$ \ follows the same as above way.$\triangleright $
\end{proof}

It is easy now, due to the symmetry between pairs of functional subspaces ${%
\mathcal{H}}_{0}^{\circledast }\times {\mathcal{H}}_{0}$ and ${\mathcal{%
\tilde{H}}}_{0}^{\circledast }\times \tilde{\mathcal{H}}_{0},$ to construct
the inverse operators to (3.4) and (3.5):
\begin{equation}
\begin{array}{l}
\mathbf{\Omega }^{-1}:=1-{\psi }\Omega _{(x_{0},t_{0})}^{-1}[\tilde{\varphi},%
\tilde{\psi}]\int_{{\mathcal{S}}(\sigma _{(x,t)}^{(m-1)},\sigma
_{(x_{0},t_{0})}^{(m-1)})}\tilde{Z}^{(m)}[\tilde{\varphi},\cdot ], \\
\mathbf{\Omega }^{\circledast ,-1}:=1-{\varphi (}\tilde{\Omega}%
_{(x_{0},t_{0})}^{^{\circledast }}[\tilde{\varphi},\tilde{\psi}])^{-1}\int_{{%
\mathcal{S}}(\sigma _{(x,t)}^{(m-1)},\sigma _{(x_{0},t_{0})}^{(m-1)})}%
\overline{\tilde{Z}}^{(m),\intercal }[\cdot ,\tilde{\psi}],%
\end{array}
\label{3.7}
\end{equation}%
where by definition,
\begin{equation*}
\lbrack <\tilde{\varphi},\tilde{L}\tilde{\psi}-\frac{\partial \tilde{\psi}}{%
\partial t}>-<\tilde{L}^{\ast }\tilde{\varphi}+\frac{\partial \tilde{\varphi}%
}{\partial t},\tilde{\psi}>]dt\wedge dx=d\tilde{Z}^{(m)}[\tilde{\varphi},%
\tilde{\psi}],
\end{equation*}%
$\tilde{Z}^{(m)}[\tilde{\varphi},\tilde{\psi}]:=d\tilde{\Omega}^{(m-1)}[%
\tilde{\varphi},\tilde{\psi}]\in \Lambda (\mathbb{R}^{m+1};\mathbb{C}),$ $(%
\tilde{\varphi},\tilde{\psi})\in \tilde{\mathcal{H}}_{0}^{^{\circledast
}}\times \tilde{\mathcal{H}}_{0}$ and the pair of \ functions $(\varphi
,\psi )\in {\mathcal{H}}_{0}^{^{\circledast }}\times {\mathcal{H}}_{0}$
satisfies the necessary inverse mappings conditions:
\begin{equation}
\psi =\mathbf{\Omega }^{-1}(\tilde{\psi}),\qquad \varphi =\mathbf{\Omega }%
^{\circledast ,-1}(\tilde{\varphi}),  \label{3.8}
\end{equation}%
which can be checked easily by simple calculations.

For the construction of the Delsarte transformed operator $\tilde{L}:{%
\mathcal{H}}\rightarrow {\mathcal{H}}$ to be finished, it is necessary to
state that this operator is differential too. The following theorem holds.

\begin{theorem}
The Delsarte transformed operator
$\frac{\partial}{\partial t}-\tilde{L}=\hat{\Omega}\left(\frac{\partial}{\partial t}-L\right)\hat{\Omega}^{-1}
:{\mathcal{H}} \rightarrow {\mathcal{H}}$ is purely differential on the whole space $\mathcal{H}$
for any suitably chosen hypersurface ${\mathcal{S}}(\sigma_{(x,t)}^{(n-1)},\sigma_{(x_{0},t_{0})}^{(n-1)})
\subset {l} \times{R^{n}}.$
\end{theorem}

For proving the theorem one needs to show that the formal
pseudo-differential expression corresponding to the operator $\tilde{L}:{%
\mathcal{H}}\rightarrow {\mathcal{H}}$ defined by (3.5) contains no integral
element. Making use of an idea devised in [5,10], one can formulate such a
lemma.

\begin{lemma}
A multidimensional pseudo-differential operator
$L:L_{2,-}(R^{m};C^{N})\rightarrow L_{2,-}(R^{m};C^{N})$ is
purely differential iff the following equality
\begin{equation}
\left( \left< h,\left( L \frac{\partial^{|\alpha|}}{\partial x^{\alpha}}\right)_{+}f \right> \right)=
\left( \left< h,L_{+} \frac{\partial^{|\alpha|}}{\partial x^{\alpha}}f \right> \right)
\end{equation}
holds for any $|\alpha| \in Z_{+}$ and all $(h,f) \in L_{2,-}(R^{m};C^{N}) \times L_{2,-}(R^{m}; C^{N}),$ that is the condition (3.11)
is equivalent to the equality $L_{+}=L,$ where as usually, the sign $"(...)_{+}"$
means the purely differential part of the corresponding expression inside the brackets.
\end{lemma}

\begin{proof}
(of \ Theorem 3.2.) Based on Lemma 3.3 and the exact expression (3.5) of the
reduced on $L_{2}(\mathbb{R}^{m};\mathbb{C}^{N})$ the operator $\tilde{L},$
similarly to calculations in [10], one finds right away that the reduced on $%
L_{2}(\mathbb{R}^{m};\mathbb{C}^{N})$ operator $\tilde{L},$ depending only
on a pair of homological cycles $\sigma _{(x,t)}^{(m-1)}$ and $\sigma
_{(x_{0},t_{0})}^{(m-1)}$ marked by points $(x,t)$ and $(x_{0},t_{0})\in
\mathbb{R}^{m}\times l,$ is purely differential in $L_{2}(\mathbb{R}^{m};%
\mathbb{C}^{N}),$ thereby proving the theorem.$\triangleright $
\end{proof}

It is natural to consider now a degenerate case when the operator $L:%
\mathcal{H\rightarrow H}$ doesn't depend on the evolution parameter $t\in l.$
Then one can construct closed subspace ${\mathcal{H}}_{0}\subset {\mathcal{H}%
}_{-}:=L_{2,-}(l;L_{2}(\mathbb{R}^{m};\mathbb{C}^{N}))$ \ as follows:
\begin{eqnarray}
{\mathcal{H}}_{0} &=&\{\psi \in {\mathcal{H}}_{-}:\psi (t;x|\lambda ,\xi
)=e^{\lambda t}{\psi }_{\lambda }(x;\xi ),\text{\ }{\psi }_{\lambda }\in
L_{2,0}(\mathbb{R}^{m};\mathbb{C}^{N}):  \notag \\
\text{ }{\psi }_{\lambda }|_{\Gamma } &=&0,\text{ }\lambda \in \sigma (L),%
\text{ }\xi \in \Sigma _{\sigma }\},  \label{3.10}
\end{eqnarray}%
where $\sigma (L)\subset \mathbb{C}$ \ is the generalized spectrum of the
extended operator $L:L_{2,-}(\mathbb{R}^{m};\mathbb{C}^{N})$ $\ \rightarrow
L_{2,-}(\mathbb{R}^{m};\mathbb{C}^{N})$ \ in a suitably Hilbert-Schmidt
rigged \cite{BS, Be} Hilbert space $L_{2,-}(\mathbb{R}^{m};\mathbb{C}^{N}),$
$L{\psi }_{\lambda }=\lambda {\psi }_{\lambda },$ $\Sigma _{\sigma }\subset
\Sigma $ is some subset, and $t\in l$ is considered as a parameter.
Correspondingly, the conjugated space ${\mathcal{H}}_{0}^{^{\circledast }}$
\ is defined as
\begin{eqnarray}
{\mathcal{H}}_{0}^{^{\circledast }} &=&\{\varphi \in {\mathcal{H}}^{\ast
}:\varphi (t;x|\lambda ,\xi )=e^{-\bar{\lambda}t}{\varphi }_{\lambda }(x;\xi
),\text{ \ }{\varphi }_{\lambda }\in L_{2,0}^{\circledast }(\mathbb{R}^{m};%
\mathbb{C}^{N}):  \notag \\
\text{ \ \ \ \ \ \ \ \ \ \ \ \ }{\varphi }_{\lambda }|_{\Gamma } &=&0,\bar{%
\lambda}\in \sigma (L^{\ast }),\text{ }\xi \in \Sigma _{\sigma }\text{ }\}.
\label{3.11}
\end{eqnarray}%
Moreover, we can here identify the $\rho $-measured set $\Sigma $ with the
product $\Sigma =(\bar{\sigma}(L^{\ast })\cap \sigma (L))\times \Sigma
_{\sigma }$ and take, correspondingly, $d\rho (\lambda ;\xi )=d\rho _{\sigma
}(\lambda )\odot d\rho _{\Sigma _{\sigma }}$ with $\lambda \in (\bar{\sigma}%
(L^{\ast })\cap \sigma (L))$ and $\xi \in \Sigma _{\sigma }.$ If now to
choose a pair of \ homologically conjugated cycles $\sigma
_{(x,t_{0})}^{(m-1)},\sigma _{(x_{0},t_{0})}^{(m-1)}$ \ lying in the space $%
\mathbb{R}^{m}$ for any $t=t_{0}\in \mathbb{R}$ \ being fixed, one easily
finds that the corresponding Delsarte transmutation operator $\mathbf{\Omega
}:{\mathcal{H}}\rightarrow {\mathcal{H}}$\ reduces to the operator $\mathbf{%
\Omega }:L_{2}(\mathbb{R}^{m};\mathbb{C}^{N})\rightarrow L_{2}(\mathbb{R}%
^{m};\mathbb{C}^{N}),$ not depending on the parameter $t\in l.$ Thus, we can
write down now, that this operator in $L_{2}(\mathbb{R}^{m};\mathbb{C}^{N})$
\ is given as follows: $\ \ \ $ \
\begin{eqnarray}
\mathbf{\Omega } &=&1-\int_{(\bar{\sigma}(L^{\ast })\cap \sigma (L))}d\rho
_{(\sigma )}(\lambda )\int_{\Sigma _{\sigma }\times \Sigma _{\sigma }}d\rho
_{(\Sigma _{\sigma })}(\xi )d\rho _{(\Sigma _{\sigma })}(\eta ){\tilde{\psi}}%
_{\lambda }(\xi )  \label{3.12} \\
&&\times (\Omega _{(x_{0},t_{0})}[\varphi _{\lambda },\psi _{\lambda
}])^{-1}(\xi ,\eta )\int_{{\mathcal{S}}(\sigma _{(x,t_{0})}^{(m-1)},\sigma
_{(x_{0},t_{0})}^{(m-1)})}Z^{(m)}[\varphi _{\lambda },\cdot ](\eta )  \notag
\end{eqnarray}%
and, correspondingly, the operator $\mathbf{\Omega }^{^{\circledast
}}:L_{2}^{\ast }(\mathbb{R}^{m};\mathbb{C}^{N})\rightarrow L_{2}^{\ast }(%
\mathbb{R}^{m};\mathbb{C}^{N})$ is given as
\begin{eqnarray}
\mathbf{\Omega }^{^{\circledast }} &=&1-\int_{(\bar{\sigma}(L^{\ast })\cap
\sigma (L))}d\rho _{(\sigma )}(\lambda )\int_{\Sigma _{\sigma }\times \Sigma
_{\sigma }}d\rho _{(\Sigma _{\sigma })}(\xi )d\rho _{(\Sigma _{\sigma
})}(\eta )\tilde{\varphi}_{\lambda }(x;\xi )  \label{3.13} \\
&&\times (\mathbf{\Omega }_{(x_{0},t_{0})}^{\ast }[\varphi _{\lambda },\psi
_{\lambda }])^{-1}(\xi ,\eta )\int_{{\mathcal{S}}(\sigma
_{(x,t_{0})}^{(m-1)},\sigma _{(x_{0},t_{0})}^{(m-1)})}^{\text{ \ \ }}\bar{Z}%
^{(m),\intercal }[\cdot ,\psi _{\lambda }](\eta )  \notag
\end{eqnarray}%
where $(\varphi _{\lambda },\psi _{\nu })\in L_{2,0}^{\circledast }(\mathbb{R%
}^{m};\mathbb{C}^{N})\times L_{2,0}(\mathbb{R}^{m};\mathbb{C}^{N})$ are
generalized eigenfunctions with the generalized eigenvalues $\lambda ,\nu
\in \bar{\sigma}(L^{\ast })\cap \sigma (L)$ of the corresponding pair of
operators $L^{\ast }$ $:L_{2,-}^{\ast }(\mathbb{R}^{m};\mathbb{C}%
^{N})\rightarrow L_{2,-}^{\ast }(\mathbb{R}^{m};\mathbb{C}^{N})$ \ and $%
L:L_{2,-}(\mathbb{R}^{m};\mathbb{C}^{N})\rightarrow L_{2,-}(\mathbb{R}^{m};%
\mathbb{C}^{N}).$ Since the differential $dt=0$ in the case (3.12) and
(3.13), for the differential $m$-form $Z^{(m)}[\varphi _{\lambda },\psi
_{\nu }]\in \Lambda ^{m}(\mathbb{R}^{m};\mathbb{C})$ one gets the simple
expression
\begin{equation}
Z^{(m)}[\varphi _{\lambda },\psi _{\nu }](\xi ,\eta )=-dx{\bar{\varphi}}%
_{\lambda }^{\intercal }(x;\xi ){\psi }_{\nu }(x;\eta )  \label{3.14}
\end{equation}%
with \ $\lambda ,\nu \in \bar{\sigma}(L^{\ast })\cap \sigma (L)$ and $(\xi
,\eta )\in \Sigma _{\sigma }\times \Sigma _{\sigma }.$ Thus the
corresponding operator (3.12) in $L_{2}(\mathbb{R}^{m};\mathbb{C}^{N})$ \
takes the form%
\begin{equation}
\mathbf{\Omega }=1+\int_{{\mathcal{S}}(\sigma _{(x,t_{0})}^{(m-1)},\sigma
_{(x_{0},t_{0})}^{(m-1)})}dyK(x;y)(\cdot ),  \label{3.15}
\end{equation}%
where for a fixed set of functions $({\varphi }_{\lambda },{\psi }_{\lambda
})\in L_{2,0}^{\circledast }(\mathbb{R}^{m};\mathbb{C}^{N})\times L_{2,0}(%
\mathbb{R}^{m};\mathbb{C}^{N}),$ $\lambda \in \bar{\sigma}(L^{\ast })\cap
\sigma (L),$ the kernel $K(x;y),$ $x,y\in \mathbb{R}^{m},$ is given as
follows:
\begin{eqnarray}
K(x,y) &=&-\int_{\bar{\sigma}(L^{\ast })\cap \sigma (L)}d\rho _{(\sigma
)}(\lambda )\dint\limits_{\Sigma _{\sigma }\times \Sigma _{\sigma }}d\rho
_{(\Sigma _{\sigma })}(\xi )d\rho _{(\Sigma _{\sigma })}(\eta ){\tilde{\psi}}%
_{\lambda }(x;\xi )  \label{3.16} \\
&&\times \Omega _{(x_{0},t_{0})}^{-1}[\varphi _{\lambda },\psi _{\lambda
}](\xi ,\eta ){\bar{\varphi}}_{\lambda }^{\intercal }(y;\eta ),  \notag
\end{eqnarray}%
being, evidently, of Volterra type and completely similar to that obtained
in \cite{Be} in the case of \ selfadjoint operators $L^{\ast }=L$ in a
Hilbert-Schmidt rigged Hilbert space $L_{2}(\mathbb{R}^{m};\mathbb{C}^{N}).$
The constant operator $\Omega _{(x_{0},t_{0})}[\varphi _{\lambda },\psi
_{\lambda }]:L_{2}^{\rho }(\Sigma _{\sigma };\mathbb{C})\rightarrow
L_{2}^{\rho }(\Sigma _{\sigma };\mathbb{C}),$ is defined naturally by the
topological structure of the homological hypercycle $\sigma
_{(x_{0},t_{0})}^{(m-1)}\subset \mathbb{R}^{m},$ in particular, by
asymptotic properties of the generalized eigenfunctions ${\varphi }_{\lambda
}\in L_{2,0}^{\circledast }(\mathbb{R}^{m};\mathbb{C}^{N})$ and ${\psi }%
_{\lambda }\in L_{2,0}(\mathbb{R}^{m};\mathbb{C}^{N}),$ $\lambda \in \bar{%
\sigma}(L^{\ast })\cap \sigma (L),$ as $|x|\rightarrow \infty .$ Another
useful equation on the kerenel (3.16) based only on its form looks as
follows:%
\begin{equation}
\overset{\_}{\tilde{L}}_{(x)}\text{ \ }\bar{K}(x,y)=(L_{(y)}^{\ast }\bar{K}%
^{\intercal }(x,y))^{\intercal }  \label{3.17}
\end{equation}%
for all $x,y\in \mathbb{R}^{m}.$ It is completely analogous to the equations
which were before derived in the one- and two-dimensional cases in \cite{Be}
and [3-5].

\section{Applications to spectral and soliton theories: a short sketch.}

Take a differential operator $L:L_{2}(\mathbb{R}^{m};\mathbb{C}%
^{N})\rightarrow L_{2}(\mathbb{R}^{m};\mathbb{C}^{N})$ like (1.2) and
construct its Delsarte transformation $\tilde{L}:L_{2}(\mathbb{R}^{m};%
\mathbb{C}^{N})\rightarrow L_{2}(\mathbb{R}^{m};\mathbb{C}^{N})$ \ via the
expression
\begin{equation}
\tilde{L}=\mathbf{\Omega }L\mathbf{\Omega }^{-1},  \label{4.1}
\end{equation}%
being of the same form a differential operator in $L_{2}(\mathbb{R}^{m};%
\mathbb{C}^{N}).$ Assuming that the spectral properties of the operator $L$
\ are known and simpler, one can try to study the corresponding spectral
properties of the operator $\tilde{L},$ being more complicated than $L.$
Under such transformations, as is well known, the spectrum of the operator $%
\tilde{L}$ can change significally, for instance, the discrete spectrum of $%
\tilde{L}$ can appear, leaving the essential continuous spectrum $\sigma
_{c}(\tilde{L})$ of the transformed operator $\tilde{L}$ unchangeable. An
approach realizing in part this idea was before developed in [4,5] for the
case of one and two-dimensional Dirac and Laplace operators.

Subject to soliton theory, it is necessary to take two a priori commuting
differential operators $(\frac{\partial }{\partial t}-L)$ and $(\frac{%
\partial }{\partial y}-M):{\mathcal{H}}\rightarrow {\mathcal{H}}$ with ${%
\mathcal{H}}\subset L_{2}(\mathbb{R}^{2};L_{2}(\mathbb{R}^{m};\mathbb{C}%
^{N})),$ that is
\begin{equation}
\lbrack \frac{\partial }{\partial t}-L,\frac{\partial }{\partial y}-M]=0.
\label{4.2}
\end{equation}%
Making use of a fixed Delsarte transmutation constructed for these two
operators by means of an invertible operator mapping like (3.12), (3.13),
one gets two differential operators $\frac{\partial }{\partial t}-\tilde{L}$
and $\frac{\partial }{\partial y}-\tilde{M}:{\mathcal{H}}\rightarrow {%
\mathcal{H}},$ generated by closed subspaces $\mathcal{H}_{0}^{^{\circledast
}}$ and $\mathcal{H}_{0},$ where, by definition,
\begin{eqnarray*}
\mathcal{H}_{0} &:&=\underset{L_{2,-}(\mathbb{R}^{m};\mathbb{C}^{N})}{\text{%
closure}}\{\underset{\mathbb{C}}{span}\{\psi \in \mathcal{H}_{-}:\partial
\psi |/\partial t=L\psi ,\text{ }\psi |_{t=t_{0}}=\psi _{\lambda }\in
L_{2,0}(\mathbb{R}^{m};\mathbb{C}^{N}), \\
\text{ }L\psi _{\lambda } &=&\lambda \psi _{\lambda },\text{ \ \ }\psi
_{\lambda }|_{\Gamma }=0,\text{ \ }\lambda \in \bar{\sigma}(L^{\ast })\cap
\sigma (L)\},
\end{eqnarray*}%
\begin{eqnarray*}
\mathcal{H}_{0}^{^{\circledast }} &:&=\underset{L_{2,-}(\mathbb{R}^{m};%
\mathbb{C}^{N})}{\text{closure}}\{\underset{\mathbb{C}}{span}\{\varphi \in
\mathcal{H}_{-}^{\ast }:-\partial \varphi |/\partial t=L^{\ast }\varphi ,%
\text{ }\varphi |_{t=t_{0}}=\varphi _{\lambda }\in L_{2,0}^{\circledast }(%
\mathbb{R}^{m};\mathbb{C}^{N}), \\
L\varphi _{\lambda } &=&\bar{\lambda}\varphi _{\lambda },\text{ \ \ }\varphi
_{\lambda }|_{\Gamma }=0,\text{ }\bar{\lambda}\in \bar{\sigma}(L^{\ast
})\cap \sigma (L)\},
\end{eqnarray*}%
also commuting in $\mathcal{H},$ that is
\begin{equation}
\lbrack \frac{\partial }{\partial t}-\tilde{L},\frac{\partial }{\partial y}-%
\tilde{M}]=0.  \label{4.3}
\end{equation}%
The latter, so called a Zakharov-Shabat operator equality in ${\mathcal{H}},$
is as well known [7,8], equivalent to some system of compatible nonlinear
evolution equations upon the coefficients of the operators $\tilde{L}$ and $%
\tilde{M}.$

Moreover, since flows $\frac{\partial }{\partial t}$ and $\frac{\partial }{%
\partial y}$ in ${\mathcal{H}}$ are commuting, the corresponding
differential $m$-form $Z^{(m)}[\varphi ,\psi ],$ given by (2.4) and defining
the Delsarte transmutation operator $\mathbf{\Omega }:{\mathcal{H}}%
\rightarrow {\mathcal{H}},$ given by (3.12), has to be naturally changed by
a similar extended differential $m$-form $Z^{(m)}[\varphi ,\psi ],$ given by
the expression
\begin{equation}
\begin{array}{l}
Z^{(m)}[\varphi ,\psi ]=\sum_{i=\overline{1,m}}dt\wedge dx_{1}\wedge
dx_{2}\wedge ...dx_{i-1}Z_{i}^{(L)}[\varphi ,\psi ]\wedge dx_{i+1}\wedge
...\wedge dx_{m} \\
+\sum_{i=\overline{1,m}}dy\wedge dx_{1}\wedge dx_{2}\wedge
...dx_{i-1}Z_{i}^{(M)}[\varphi ,\psi ]\wedge dx_{i+1}\wedge ...\wedge dx_{m}
\\
+\varphi ^{\intercal }(x;t,y)\psi (x;t,y)dx%
\end{array}
\label{4.4}
\end{equation}%
for any pair $(\varphi ,\psi )\in \mathcal{H}_{0}^{^{\circledast }}\times
\mathcal{H}_{0}.$ It is easyly seen that the extended differential $(m+1)$%
-form $dZ^{(m)}[\varphi ,\psi ]=0$ upon the space $\mathcal{H}%
_{0}^{^{\circledast }}\times \mathcal{H}_{0},$ that is due to the Stokes
theorem [9] there exists a differential $(m-1)$-form $\Omega
^{(m-1)}[\varphi ,\psi ]\in \Lambda ^{m-1}(\mathbb{R}^{2}\times \mathbb{R}%
^{m};\mathbb{C}),$ such that
\begin{equation}
d\Omega ^{(m-1)}[\varphi ,\psi ]=Z^{(m)}[\varphi ,\psi ]  \label{4.5}
\end{equation}%
for all $(\varphi ,\psi )\in \mathcal{H}_{0}^{^{\circledast }}\times
\mathcal{H}_{0}.$ Making use of this $(m-1)$-form $\Omega
^{^{(m-1)}}[\varphi ,\psi ]\in \Lambda ^{m-1}(\mathbb{R}^{2}\times \mathbb{R}%
^{m};\mathbb{C})$ one can, similarly the way used before, construct the
corresponding invertible Delsarte transmutation operators $\mathbf{\Omega }:{%
\mathcal{H}}\rightarrow {\mathcal{H}}$ and $\ \mathbf{\Omega }%
^{^{\circledast }}:{\mathcal{H}}^{\ast }\rightarrow {\mathcal{H}}^{\ast }$
in the form like (3.12) and (3.13), but depending on hyper-surface ${%
\mathcal{S}}(\sigma _{(x;t,y)}^{(m-1)},\sigma
_{(x_{0};t_{0},y_{0})}^{(m-1)}) $\ $\subset \mathbb{R}^{2}\times \mathbb{R}%
^{m},$ spanned between two $(m-1) $-dimensional homologically conjugated
cycles $\sigma _{(x;t,y)}^{(m-1)},\sigma _{(x;t_{0},y_{0})}^{(m-1)}$ $%
\subset \mathbb{R}^{2}\times \mathbb{R}^{m}.$

This construction finishes our discussion of Delsarte transmutation
operators for a commuting pair of operators $\frac{\partial }{\partial t}-L$
and $\frac{\partial }{\partial y}-M$ acting in a parametrically dependent
functional space $\mathcal{H}.$ In the case of the measure $\rho $ on $%
\Sigma $ chosen discrete, the corresponding Delsarte transmutation operators
is often called a Darboux-Backlund transformation \cite{MS, PS} of a given
pair of operators $\frac{\partial }{\partial t}-L$ and $\frac{\partial }{%
\partial y}-M,$ giving rise to the Darboux type formulas like (3.2) and
operator equalities
\begin{equation}
\tilde{L}=L-[\mathbf{\Omega },\frac{\partial }{\partial t}-L]\mathbf{\Omega }%
^{-1},\text{ \ }\tilde{M}=M-[\mathbf{\Omega },\frac{\partial }{\partial y}-M]%
\mathbf{\Omega }^{-1},  \label{4.6}
\end{equation}%
giving rise to the corresponding Backlund type expressions for the
coefficients of the Delsarte transformed operators $\tilde{L}$ and $\tilde{M}
$ \ in $\mathcal{H}.$ The latter, as well known, is of great importance for
finding new soliton like solutions to the system of evolution equations,
equivalent to the operator equality (4.3). Some applications of this
algorithm to finding exact solutions of the Davey-Stuartson and
Nizhnik-Novikov-Veselov equations are done, for instance, in [5,9]. And the
last note concerns the applications of the theory devised above to finding
the corresponding Delsarte transmutation operators for multidimensional
matrix differential operator pencils rationally depending on a
\textquotedblright spectral\textquotedblright\ parameter $\lambda \in
\mathbb{C}:$ \ this case can be treated similarly to that considered above
making use inside the operators $\partial /\partial t-L$ and $\partial
/\partial y-M,$ taken in the form%
\begin{eqnarray}
\partial /\partial t-L &:&=\partial /\partial t-\sum_{|\alpha
|=0}^{n(L)}a_{\alpha }(t;x|\lambda )\frac{\partial ^{|\alpha |}}{\partial
x^{\alpha }},  \label{4.7} \\
\partial /\partial y-M &:&=\partial /\partial y-\sum_{|\beta
|=0}^{n(M)}b_{\beta }(t;x|\lambda )\frac{\partial ^{|\beta |}}{\partial
x^{\beta }},  \notag
\end{eqnarray}%
where $a_{\alpha },b_{\beta }\in C^{1}(\mathbb{R}_{(t,y)}^{2};\mathcal{S}(%
\mathbb{R}^{m};End\mathbb{C}^{N}))\otimes \mathbb{C}_{\lambda }$ for all $%
|\alpha |=\overline{0,n(L)},|\beta |=\overline{0,n(M)},$ $\ n(L),n(M)\in
\mathbb{Z}_{+},$ of the change of the variable $\lambda \in \mathbb{C}$ by
the operation of differentiation $\partial /\partial \tau ,$ $\tau \in
\mathbb{R},$ and next applying the developed before approach to constructing
the corresponding Delsarte transmutation operators in the functional space $%
C^{1}(\mathbb{R}_{\tau }\mathbb{\times }\mathbb{R}_{(t,y)}^{2};L_{2}(\mathbb{%
R}^{m};H)),$ and at the end returning back to the starting picture putting,
correspondingly, the closed subspaces
\begin{eqnarray*}
{\mathcal{H}}_{0} &=&\{\psi \in {\mathcal{H}}:\psi (\tau ;x;y,t|\lambda ;\xi
)=e^{\lambda \tau }{\psi }_{\lambda }(x;y,t;\xi ), \\
{\psi }_{\lambda } &\in &L_{2}(\mathbb{R}^{2};L_{2,0}(\mathbb{R}^{m};\mathbb{%
C}^{N})),\text{ }{\psi }_{\lambda }(x,y;\xi )|_{\Gamma }=0,\text{ }\xi \in
\Sigma _{\sigma },\lambda \in \sigma (L)\text{ }\},
\end{eqnarray*}%
\begin{eqnarray*}
{\mathcal{H}}_{0}^{^{\circledast }} &=&\{\varphi \in {\mathcal{H}}^{\ast
}:\varphi (\tau ;x;t|\lambda ;\xi )=e^{-\bar{\lambda}t}{\varphi }_{\lambda
}(x;y;\xi ), \\
{\varphi }_{\lambda } &\in &L_{2}(\mathbb{R}^{2};L_{2,0}^{\circledast }(%
\mathbb{R}^{m};\mathbb{C}^{N})),\text{ }{\varphi }_{\lambda }|_{\Gamma }=0,%
\text{ }\xi \in \Sigma _{\sigma },\bar{\lambda}\in \sigma (L^{\ast }),
\end{eqnarray*}%
thereby getting the corresponding two conjucated Delsarte transmutation
operators like (\ref{3.1}), acting now in the spaces $L_{2}(\mathbb{R}^{m};%
\mathbb{C}^{N})$ and $L_{2}^{\ast }(\mathbb{R}^{m};\mathbb{C}^{N}),$
correspondingly. On these aspects of this technique and on its applications
we plan to stop in more detail in another palce.

\section{Acknowledgements}

Authors are cordially thankful to prof. Nizhnik L.P. (Kyiv, Inst. of Math.at
NAS), prof. T. Winiarska (Krakow, PK), profs. A. Pelczar and J. Ombach
(Krakow, UJ), prof. St. Brzychczy ( Krakow, AGH) and prof. Z. Peradzynski
(Warszawa, UW) for valuable discussions during their seminars of some
aspects of problems studied in the work.

\end{document}